\documentclass[conference]{IEEEtran}
\IEEEoverridecommandlockouts
\usepackage{cite}
\usepackage{amsmath,amssymb,amsfonts}
\usepackage{upgreek}
\usepackage{graphicx}
\usepackage{subfigure}
\usepackage{textcomp}
\usepackage{xcolor}
\usepackage{algorithm}
\usepackage[noend]{algpseudocode}
\usepackage{amsmath}
\usepackage[caption=false,font=normalsize,labelfont=sf,textfont=sf]{subfig}
\usepackage{stfloats}
\usepackage[linewidth=1pt]{mdframed}
\usepackage[mathscr]{eucal}
\usepackage{balance}

\newtheorem{corollary}{Corollary}
\def\BibTeX{{\rm B\kern-.05em{\sc i\kern-.025em b}\kern-.08em
    T\kern-.1667em\lower.7ex\hbox{E}\kern-.125emX}}

\begin{document}

\title{Uplink Power Control for Extremely Large-Scale MIMO with Multi-Agent Reinforcement Learning and Fuzzy Logic\\
}

\author{\IEEEauthorblockN{Ziheng Liu}
\IEEEauthorblockA{\textit{School of Electronic and Information} \\
\textit{Engineering}\\
\textit{Beijing Jiaotong University}\\
Beijing, China \\
zihengliu01@163.com}
\and
\IEEEauthorblockN{Zhilong Liu}
\IEEEauthorblockA{\textit{School of Electronic and Information} \\
\textit{Engineering}\\
\textit{Beijing Jiaotong University}\\
Beijing, China \\
zhilongliu@bjtu.edu.cn}
\and
\IEEEauthorblockN{Jiayi Zhang}
\IEEEauthorblockA{\textit{School of Electronic and Information} \\
\textit{Engineering}\\
\textit{Beijing Jiaotong University}\\
Beijing, China \\
zhangjiayi@bjtu.edu.cn}
\and
\IEEEauthorblockN{Huahua Xiao}
\IEEEauthorblockA{\textit{State Key Laboratory of Mobile Network} \\
\textit{and Mobile Multimedia Technology},\\
\textit{ZTE Corporation}\\
Beijing, China \\
xiao.huahua@zte.com.cn}
\and
\IEEEauthorblockN{Bo Ai}
\IEEEauthorblockA{\textit{State Key Laboratory of Rail Traffic} \\
\textit{Control and Safety}\\
\textit{Beijing Jiaotong University}\\
Beijing, China \\
boai@bjtu.edu.cn}
\and
\IEEEauthorblockN{Derrick Wing Kwan Ng}
\IEEEauthorblockA{\textit{School of Electrical Engineering and} \\
\textit{Telecommunications}\\
\textit{University of New South Wales}\\
Sydney, Australia \\
w.k.ng@unsw.edu.au}
}
\maketitle

\begin{abstract}
In this paper, we investigate the uplink transmit power optimization problem in cell-free (CF) extremely large-scale multiple-input multiple-output (XL-MIMO) systems. Instead of applying the traditional methods, we propose two signal processing architectures: the centralized training and centralized execution with fuzzy logic as well as the centralized training and decentralized execution with fuzzy logic, respectively, which adopt the amalgamation of multi-agent reinforcement learning (MARL) and fuzzy logic to solve the design problem of power control for the maximization of the system spectral efficiency (SE). Furthermore, the uplink performance of the system adopting maximum ratio (MR) combining and local minimum mean-squared error (L-MMSE) combining is evaluated. Our results show that the proposed methods with fuzzy logic outperform the conventional MARL-based method and signal processing methods in terms of computational complexity. Also, the SE performance under MR combining is even better than that of the conventional MARL-based method.
\end{abstract}


\section{Introduction}
To cope with the rapid growth of data throughput in wireless communication networks, various new communication paradigms and new technologies have been proposed to satisfy the increasing demand of communication quality. In particular, extremely large-scale multiple-input multiple-output (XL-MIMO) is regarded as a promising technology to provide higher spectral efficiency (SE) and energy efficiency (EE) for the next-generation wireless communication systems \cite{[1],[2],[3]}. Compared with the conventional cell-free (CF) massive MIMO (mMIMO) \cite{[4],[12],[14]}, the novel XL-MIMO deploys as many antennas as possible in a compact space that leads to a fundamental change of paradigm in the electromagnetic (EM) characteristics \cite{[1]}. Especially, the commonly adopted original uniform plane wave (UPW) model based on far-field assumption is not valid in the XL-MIMO, as near-field propagation usually dominates in the latter case.

Specifically, most of the research on XL-MIMO systems have shifted the focus from the far-field characteristics to the near-field characteristics. For instance, in \cite{[2]}, the characteristics of near-field radiation adopting to the spherical waveform propagation model were presented. Moreover, the authors in \cite{[3]} proposed an efficient hybrid-field channel estimation scheme, revealing the channel feature of XL-MIMO systems.

With the limited communication resources and inter-user interference, designing a proper power control method is necessary to unlock the potential of XL-MIMO and to optimize the system performance. Indeed, the conventional power control methods have been well-studied in the literature \cite{[5],[6]}, which can achieve excellent performance. However, these methods require high computational complexity, which are not applicable to the practical implementation of XL-MIMO.

Meanwhile, multi-agent reinforcement learning (MARL), a disruptive technique has been adopted in numerous domains, such as autonomous driving and robotics \cite{[7]}. In particular, MARL can improve the overall learning performance and achieve its objective through interactions. Recently, MARL-based solutions have been studied for large-scale CF mMIMO systems. For instance, in \cite{[8]}, a MARL-based optimization of pilot assignment for mitigating pilot contamination was proposed, which can effectively reduce the computational complexity. Also, in \cite{[9]}, the joint communication and computing resource allocation problem was solved by the fully distributed MARL-based method.
Unfortunately, for large-scale MARL, the joint learning is unlikely to be implemented in practical application scenarios due to its high computational complexity. Furthermore, the associated independent learning cannot guarantee the convergence of results. As such, the authors in \cite{[10]} leveraged fuzzy logic to deal with the challenges mentioned above via designing a new paradigm for MARL.

Motivated by the application of fuzzy logic for large-scale MARL in \cite{[10]}, this paper introduces a novel MARL-based uplink power control method for CF XL-MIMO. The major contributions of this paper are as follows:

\begin{itemize}
\item We first develop two different processing schemes with fuzzy logic for CF XL-MIMO systems, i.e., centralized training and centralized execution with fuzzy logic (FL-CTCE), centralized training and decentralized execution with fuzzy logic (FL-CTDE), respectively.

\item We leverage the proposed methods to optimize the SE performance with uplink power control, which can approach the performance achieved by applying the convex optimization solver but with reduced computational complexity compared with the latter.

\end{itemize}

\section{System Model}
\begin{figure}[t]
    \centering
    \includegraphics[width=2.95in]{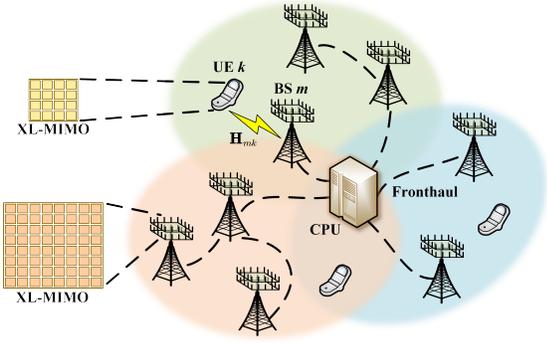}
    \caption{Illustration of a CF XL-MIMO system.\qquad\qquad\qquad\qquad\qquad\qquad}
    \label{fig1}
\end{figure}
\newcounter{mytempeqncnt_1}
\begin{figure*}[b]
\hrulefill
\normalsize
\setcounter{mytempeqncnt_1}{\value{equation}}
\setcounter{equation}{2}
\begin{equation}
\begin{split}
\mathbf{H}_{mk}&=\sqrt{{N_r}{N_s}}\sum_{(\ell_x,\ell_y) \in \varepsilon_r}\sum_{(m_x,m_y) \in \varepsilon_s}H_a^{(mk)}(\ell_x,\ell_y,m_x,m_y)\mathbf{a}_r(\ell_x,\ell_y,\mathbf{r}^{(m)})\mathbf{a}_s(m_x,m_y,\mathbf{s}^{(k)}).
\end{split}
\end{equation}
\setcounter{equation}{\value{mytempeqncnt_1}}
\label{eq3}
\end{figure*}
This paper considers a CF XL-MIMO network with $M$ BSs and $K$ UEs that are arbitrarily distributed in a large service area, where the BSs are connected to a central processing unit (CPU) via perfect fronthaul links \cite{[18]}, as illustrated in Fig. \ref{fig1}. Each BS is comprised of a planar XL-MIMO with $N_r = {N_{V_r}}{N_{H_r}}$ patch antennas \cite{[11]}, where the patch antennas spacing $\Delta_r$ is less than half of the carrier wavelength, $\lambda/2$. The antennas at each BS are indexed row-by-row by $n \in [1,N_r]$, and the location of BS $m$ with respect to the origin is $\mathbf{r}_m = [r_{m,x},r_{m,y},r_{m,z}]^T$. The receive vector at the $M$ BSs can be denoted as
$\mathbf{a}_{r}(\mathbf{k},\mathbf{r})=[\mathbf{a}_{r,1}(\mathbf{k},\mathbf{r}), \ldots, \mathbf{a}_{r,M}(\mathbf{k},\mathbf{r})]$ with
\begin{equation}
\begin{split}
\mathbf{a}_{r,m}(\mathbf{k},\mathbf{r})=[e^{j\mathbf{k}_{r,m}(\varphi,\theta)^T{\mathbf{r}_m^{(1)}}}, \ldots, e^{j\mathbf{k}_{r,m}(\varphi,\theta)^T{\mathbf{r}_{m}^{(N_r)}}}]^T,
\end{split}
\label{eq1}
\end{equation}
where $\mathbf{k}_{r,m}(\varphi,\theta)=k[\cos(\theta)\cos(\varphi),\cos(\theta)\sin(\varphi),\sin(\theta)] \in \mathbb{R}^3$
is the receive wave vector with the receive elevation angle $\theta$ and the receive azimuth angle $\varphi$ at BS $m$, $\forall m \in \{1, \ldots, M\}$.

Similarly, each UE is equipped with $N_s$ patch antennas with the spacing $\Delta_s$, and the location of UE $k$ is denoted by
$\mathbf{s}_k = [s_{k,x},s_{k,y},s_{k,z}]^T$.
The transmit signals from all the $K$ UEs can be denoted as
$\mathbf{a}_{s}(\boldsymbol{\kappa},\mathbf{s})=[\mathbf{a}_{s,1}(\boldsymbol{\kappa},\mathbf{s}), \ldots, \mathbf{a}_{s,K}(\boldsymbol{\kappa},\mathbf{s})]$,
and the transmitted signal from UE $k$ is
\begin{equation}
\begin{split}
\mathbf{a}_{s,k}(\boldsymbol{\kappa},\mathbf{s})=[e^{j\boldsymbol{\kappa}_{s,k}(\varphi,\theta)^T{\mathbf{s}_k^{(1)}}}, \ldots, e^{j\boldsymbol{\kappa}_{s,k}(\varphi,\theta)^T{\mathbf{s}_k^{(N_s)}}}]^T,
\end{split}
\label{eq2}
\end{equation}
where $\boldsymbol{\kappa}_{s,k}(\varphi,\theta)=k[\cos(\theta)\cos(\varphi),\cos(\theta)\sin(\varphi),\sin(\theta)] \in \mathbb{R}^3$ is the transmit wave vector at UE $k$, $\forall k \in \{1, \ldots, K\}$.

\subsection{Channel Model}
Based on the multi-BS multi-UE system considered above, the channel coefficient between BS $m$ and UE $k$ is modeled as
$\mathbf{G}_{mk} = \sqrt{\beta_{mk}}\mathbf{H}_{mk}$, where $\sqrt{\beta_{mk}}$ represents the large-scale fading (LSF) coefficient and $\mathbf{H}_{mk}$ represents the small-scale fading coefficient, respectively. Following the single-BS multi-UE channel model proposed by the authors in \cite{[11]} for XL-MIMO, the corresponding small-scale fading coefficient $\mathbf{H}_{mk} \in \mathbb{C}^{N_r\times{N_s}}$ can be defined as (3), shown at the bottom of the page, where $H_a^{(mk)}(\ell_x,\ell_y,m_x,m_y)$ is the Fourier coefficient with variance $\sigma_{mk}^2(\ell_x,\ell_y,m_x,m_y)$, satisfying
\begin{equation}
\setcounter{equation}{4}
H_a^{(mk)}(\ell_x,\ell_y,m_x,m_y) \sim \mathcal{N}_\mathbb{C}(0,\sigma_{mk}^2(\ell_x,\ell_y,m_x,m_y)).
\label{eq4}
\end{equation}

\subsection{Uplink Data Transmission}
In CF XL-MIMO, all the UEs send the signal to all the BSs \cite{[15],[16],[17]}.
The transmitted symbol of UE $k$ is denoted by $\mathbf{x}_k = [x_{k,1}, \ldots, x_{k,N_s}]^T$, satisfying $\mathbf{x}_k = \sqrt{p_{k,N_s}}\mathbf{s}_k$ and $\text{tr}(\mathbf{x}_k\mathbf{x}_k^H) =N_s p_{k,N_s}$.
The received signals at BS $m$ is
\begin{equation}
\mathbf{y}_m=\sum_{k=1}^{K}\sqrt{p_{k,N_s}}{\mathbf{G}_{mk}}\mathbf{s}_k+\mathbf{n}_m=\sum_{k=1}^{K}{\mathbf{G}_{mk}}\mathbf{x}_k+\mathbf{n}_m,
\label{eq5}
\end{equation}
where $\mathbf{s}_k = [s_{k,1}, \ldots, s_{k,N_s}]$ and $p_{k,N_s}$ represent the signal and the transmit power of each antenna of UE $k$, respectively. Let $\mathbf{V}_{mk} \in \mathbb{C}^{{N_r}\times{N_s}}$ denote the combining matrix designed by BS $m$ for UE $k$. Then, the local estimation of the transmitted symbol $\mathbf{x}_k$ for UE $k$ at BS $m$ is
\begin{equation}
\begin{aligned}
\check{\mathbf{x}}_{mk}={\mathbf{V}_{mk}^H}\mathbf{G}_{mk}\mathbf{x}_k+\sum_{l=1,l{\neq}k}^{K}\mathbf{V}_{mk}^H\mathbf{G}_{ml}\mathbf{x}_l+{\mathbf{V}_{mk}^H}\mathbf{n}_m.
\label{eq6}
\end{aligned}
\end{equation}

Note that the large-scale fading decoding (LSFD) method requires abundant LSF parameters knowledge, which is not always feasible in CF XL-MIMO systems \cite{[13]}. Therefore, to simplify the processing, the CPU can alternatively weight the local processed signal $\check{\mathbf{x}}_{mk}$ by taking the average of them across the observations from the $M$ BSs to obtain the final symbol as (\ref{eq7}), shown at the bottom of the page. Based on the above, we can derive the uplink achievable SE as the following corollary.
\begin{figure*}[ht]
    \centering
    \includegraphics[width=6.45in]{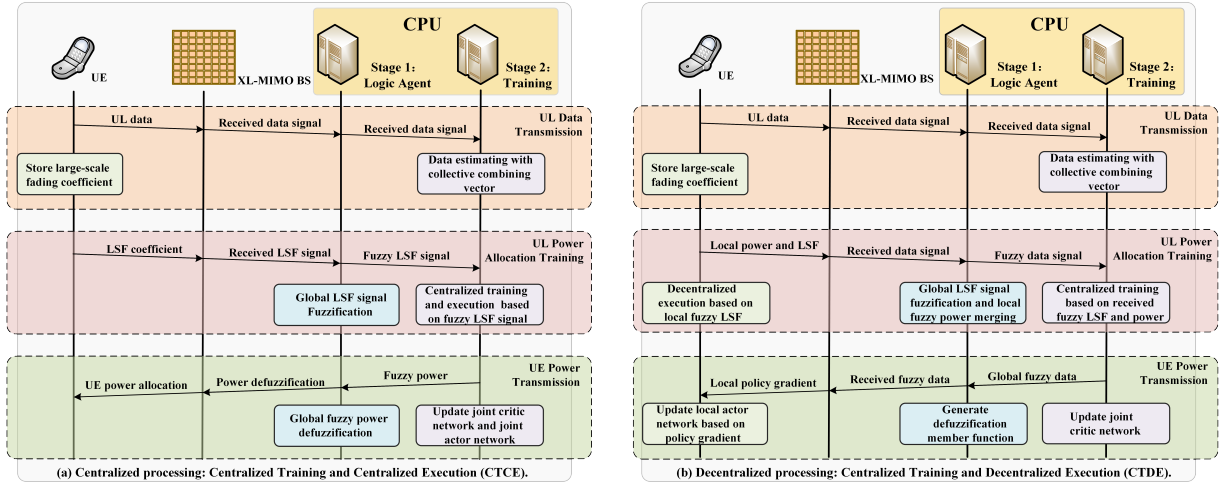}
    \caption{Different signal processing structures of FL-CTCE and FL-CTDE, respectively.\qquad\qquad\qquad\qquad\qquad\qquad\qquad\qquad\qquad\qquad\qquad\qquad\qquad\qquad\qquad\qquad\qquad\qquad}
    \label{fig2}
\end{figure*}
\begin{figure*}[hb]
\hrulefill
\begin{equation}
\begin{aligned}
\hat{\mathbf{x}}_{k}=\frac{1}{M}\sum_{m=1}^{M}\check{\mathbf{x}}_{mk}=\frac{1}{M}\sum_{m=1}^{M}{\mathbf{V}_{mk}^H}\mathbf{G}_{mk}\mathbf{x}_k+\frac{1}{M}\sum_{m=1}^{M}
\sum_{l=1,l{\neq}k}^{K}{\mathbf{V}_{mk}^H}\mathbf{G}_{ml}\mathbf{x}_l+\frac{1}{M}\sum_{m=1}^{M}{\mathbf{V}_{mk}^H}\mathbf{n}_m.
\label{eq7}
\end{aligned}
\end{equation}
\end{figure*}
\begin{corollary}
\emph{The achievable SE for UE k in the CF XL-MIMO is given by}
\begin{equation}
\begin{aligned}
\text{SE}_k = \log_{2}{\left|\mathbf{I}_{N_s}+\mathbf{E}_k^H\mathbf{\Psi}_k^{-1}\mathbf{E}_k\right|},
\label{eq8}
\end{aligned}
\end{equation}
\emph{where} $\mathbf{E}_k \triangleq \sqrt{p_{k,N_s}}\sum_{m=1}^{M}\mathbb{E}\{{\mathbf{V}_{mk}^H}\mathbf{G}_{mk}\}$
\emph{and} $\mathbf{\Psi}_k \triangleq$ $\sum_{l=1}^{K}\sum_{m=1}^{M}\sum_{m'=1}^{M}p_{l,N_s}\mathbb{E}\{{\mathbf{V}_{mk}^H}\mathbf{G}_{ml}\mathbf{V}_{m'l}^H\mathbf{G}_{m'k}\}-\mathbf{E}_k\mathbf{E}_k^H$
+ $\sum_{m=1}^{M}\mathbb{E}\{{\mathbf{V}_{mk}^H}\mathbf{n}_m\mathbf{n}_m^H\mathbf{V}_{mk}\}$.
\end{corollary}

We notice that (\ref{eq8}) are applicable along with any combining scheme matrix. One possible choice is  maximum ratio (MR) combining with $\mathbf{V}_{mk}=\mathbf{G}_{mk}$, which does not require any matrix inversion and has low computational complexity.
Besides, local minimum mean-squared error (L-MMSE) combining can also be adopted that is defined as
$\mathbf{V}_{mk}=p_{k,N_s}(\sum_{l=1}^{K}p_{l,N_s}\mathbf{G}_{ml}\mathbf{G}_{ml}^H+\sigma^2\mathbf{I}_{N_r})^{-1}\mathbf{G}_{mk}$.
Compared with the previously mentioned MR combining, although L-MMSE combining has higher computational complexity, its SE performance is far better than that of MR combining \cite{[13]}.

\section{Sum-SE Maximizing Power Control}
In this section, we formulate the uplink power control problem for the CF XL-MIMO system. The aim is to find the power allocation coefficients $\{p_{k,N_s}{:}\forall k\}$, which maximizes the sum-SE taking into account the constraints of the maximum available per UE power $P_{\max}^{ul}$.
Due to the rapid variations of the small-scale fading, it is difficult to perform instantaneous transmit power optimization. Therefore, in CF XL-MIMO, we only optimize the power allocation coefficients according to the observed LSF coefficients, and consider that each antenna has the same power. Then, the uplink power control problem can be optimized with the same power of each antenna, as follows:
\begin{equation}
\begin{aligned}
&\max_{\{p_{k,N_s}:\forall k\}} \qquad \sum_{k=1}^{K} \text{SE}_k\\
&\mbox{s.t.} \qquad N_s{p_{k,N_s}} \leq P_{\max}^{ul},\quad k = 1, \ldots, K.
\label{eq11}
\end{aligned}
\end{equation}

The uplink power control problem in (\ref{eq11}) is non-convex and the conventional optimization algorithms have high computational complexity, which makes the original solutions unfeasible in CF XL-MIMO systems. Therefore, we will propose a novel MARL-based method with fuzzy logic that overcomes the above shortcomings in the following section.

\section{MARL-Based Power Control With Fuzzy Logic}
In a multi-agent environment, each agent is composed of an actor and critic, which are adopted for action allocation and policy update, respectively. The most efficient training mechanism is the Centralized Training and Centralized Execution (CTCE), which leverages the global information to optimize policies. However, the CTCE is difficult to realize in practical scenarios due to its high computational complexity. This challenge derives the emergence of the Centralized Training and Decentralized Execution (CTDE), which simplifies centralized learning to an affordable degree.

However, in a large-scale scenarios, for the conventional MARL-based methods, they still need to be properly simplified to ensure that the designed algorithms have real-time interaction capability and scalability. Inspired by the application of fuzzy logic in \cite{[10]}, we propose a novel MARL-based uplink power control method, which leverages fuzzy logic to achieve the mapping from fuzzy agents to entities, as shown in Fig. \ref{fig2}.

\subsection{Fuzzy Logic}
This subsection introduces fuzzy logic to simplify the large-scale multi-agent system, which regards the original MARL as a fuzzy system.
In this case, we describe the proposed uplink power control problem and fuzzy logic with a MARL tuple $<{s}_{t}, {a}_{t}, {r}_{t}, \mathcal{P}, \gamma>$ at time $t$,
where ${s}_{t}=({s}_{1,t}, \ldots, {s}_{K,t})$ and ${a}_{t}=({a}_{1,t}, \ldots, {a}_{K,t})$ are the observed state and the assigned action, depending on the LSF coefficients and the uplink power allocation coefficients, respectively. ${r}_{t}$, $\mathcal{P}$ and $\gamma$ are the expected reward, the transition probability matrix and the discounted factor, respectively.
\subsubsection{Initialization}
we initialize the fuzzy state of all fuzzy agents as ${\hat{s}}_{t}=({\hat{s}}_{1,t}, \ldots, {\hat{s}}_{m,t})$, where ${\hat{s}}_{i,t}$ is randomly sampled from the observed state, and $m$ is the number of fuzzy agents.
Then, we decompose each dimension of the observation space into $m$ fuzzy sets, in which the fuzzy set for the $j$-th dimension is
$(\hat{x}_{j,t}^1, \ldots, \hat{x}_{j,t}^m)$. The corresponding membership function is $u_{\hat{x}_{j,t}^i(x)}=\exp({-\frac{1}{d_a\ast{m}}|x-\hat{x}_{j,t}^i|})$,
where $d_a$ is the dimensionality of the action space.
\subsubsection{Fuzzy action}
In a fuzzy system, we assign a policy to each fuzzy agent according to the observed fuzzy state ${\hat{s}}_{t}$, and then use defuzzification to map the fuzzy action ${\hat{a}}_{t}=({\hat{a}}_{1,t}, \ldots, {\hat{a}}_{m,t})$ to the specific action ${a}_{t}$.
Let $\mu_{k,t}^i=\prod_{i=j}^{d_a}u_{\hat{x}_{j,t}^i({x}_{j,t}^k)}$ represent the mapping relationship between $k$-th agent and $i$-th fuzzy agent \cite{[10]}. Then, the corresponding relationship can be defined as $a_{k,t} = \sum_{i=1}^{m}\bar{\mu}_{k,t}^i \times \hat{a}_{i,t}$,
where $\bar{\mu}_{k,t}^i$ is normalized mapping relationship.
\subsubsection{Fuzzy reward}
After the agents receive the specific action ${a}_{t}$, the specific reward ${r}_{t}$ can also be obtained according to the reward function.
However, because we use fuzzy agents instead of entities to interact with the environment, we need to use fuzzification to get the fuzzy reward $\hat{r}_{t}=(\hat{r}_{1,t}, \ldots, \hat{r}_{m,t})$ to complete the reinforcement learning model. Therefore, the fuzzy reward can be defined as
$\hat{r}_{i,t} = \sum_{k=1}^{K}\bar{\mu}_{k,t}^i \times {r}_{k,t}$.
\subsubsection{Fuzzy state}
However, for the fuzzy state, different from initialization, its result at time $t+1$ depends on the mapping relationship $\mu_{k,t}^i$ at time $t$ and the abstract action ${s}_{t+1}$ at time $t+1$. Therefore, for the $i$-th fuzzy agent, the state transition relationship is
$\hat{s}_{i,t+1} = \sum_{k=1}^{K}\bar{\mu}_{k,t}^i \times {s}_{k,t+1}$.

\subsection{FL-CTCE for Maximizing SE of CF XL-MIMO}
\begin{algorithm}[ht]
  \caption{FL-CTCE and FL-CTDE for Maximizing SE}
  \label{alg::conjugateGradient}
  \begin{algorithmic}[1]
      \State
        \textbf{Initialize} observations of fuzzy agents: $\hat{s}_{1,t_0}$, $\hat{s}_{2,t_0}$, \ldots, $\hat{s}_{m,t_0}$, which randomly sampled from the observations of the UE agents:
        $s_{1,t_0}$, $s_{2,t_0}$, \ldots, $s_{K,t_0}$
      \For {episode = 1 to $EP$}
        \State Evaluation-network actor determines the uplink power allocation: $\hat{a}_{i,t}$ = $\pi_i$($\hat{s}_{i,t}|i = 1, 2, \ldots, m$)
        \State Calculate the actual actions $a_{i,t}(i = 1, 2, \ldots, K)$ by defuzzification: $a_{k,t} = \sum_{i=1}^{m}\bar{\mu}_{k,t}^i \times \hat{a}_{i,t}$
        \State Obtain the actual rewards $r_{i,t}$ with reward function
        \State Calculate the fuzzy rewards $\hat{r}_{i,t}(i = 1, 2, \ldots, m)$ by fuzzification: $\hat{r}_{i,t} = \sum_{k=1}^{K}\bar{\mu}_{k,t}^i \times {r}_{k,t}$
        \State Get the next actual observations $s_{i,t+1}$ after env update
        \State Calculate the next fuzzy observations $\hat{s}_{i,t+1}(i = 1, 2, \ldots, m)$ by fuzzification: $\hat{s}_{i,t+1} = \sum_{k=1}^{K}\bar{\mu}_{k,t}^i \times {s}_{k,t+1}$
        \State Update the membership function with $u_{\hat{x}_{j,t+1}^i(\hat{x}_{j,t+1}^k)}$
        \If {FL-CTCE}
            \State Store fuzzy experience $<\hat{s}_{t}, \hat{a}_{t}, \hat{r}_{t}, \hat{s}_{t+1}>$ to the replay buffer $\mathcal{D}$
            \If {update the network}
                \State Sample a mini-batch $\mathcal{B}$ from $\mathcal{D}$ randomly
                \State Calculate the loss function of joint critic network $L(\theta_{Q_\pi})$ with the global information: equation (\ref{eq17})
                \State Update the weights of joint critic network with joint loss function $L(\theta_{Q_\pi})$
                \State Calculate the policy gradient of actor network $\Delta_{\theta_\pi}J(\theta_\pi)$ with the global information: equation (\ref{eq16})
            \EndIf
        \EndIf
        \If {FL-CTDE}
            \State Store fuzzy experience $<\hat{s}_{i,t}, \hat{a}_{i,t}, \hat{r}_{i,t}, \hat{s}_{i,t+1}>$ to the replay buffer $\mathcal{D}_i(i = 1, 2, \ldots, m)$
            \If {update the network}
                \State Sample a mini-batch $\mathcal{B}_i$ from $\mathcal{D}_i$ randomly
                \State Calculate the loss function of joint critic network $L(\theta_{Q_\pi})$ with the global information: equation (\ref{eq23})
                \State Update the weights of joint critic network with joint loss function $L(\theta_{Q_\pi})$
                \State Calculate the policy gradient of actor network $\Delta_{\theta_{\pi_i}}J(\theta_{\pi_i})$ with partial global information: equation (\ref{eq22})
            \EndIf
        \EndIf
      \EndFor
  \end{algorithmic}
\end{algorithm}
The processing flow of the FL-CTCE architecture is shown in Fig. \ref{fig2}(a), the CPU uniformly completes the action allocation and policy update. In essence, the FL-CTCE based on the deep deterministic policy gradient (DDPG) algorithm still follows the $actor$-$critic$ approach, combining the current $evaluation$ actor network ${\theta_\pi}$ and $evaluation$ critic network ${\theta_{Q_\pi}}$ with an additional $target$ actor network ${\theta_\pi'}$ and $target$ critic network ${\theta_{Q_{\pi'}}}$ for an improved convergence rate.

Besides, the policy is assigned to fuzzy agents rather than agents themselves, so that only fuzzy agents participate in the training process. The objective function for the joint policy $\pi$ is $L(\pi) = \sum_{\hat{s}_t}p_{\pi}(\hat{s}_t)\sum_{\hat{a}_t}\pi(\hat{a}_t|\hat{s}_t)\hat{r}_t$,
where $p_{\pi}(\hat{s}_t)$ and $\pi(\hat{a}_t|\hat{s}_t)$ are the stationary distribution for global abstract observation and the probability of assigned actions $\hat{a}_t$, respectively. Let $Q_\pi(\hat{s}_t,\hat{a}_t)$ represent the global action value. Additionally, the corresponding policy gradient of the joint actor network estimated by all fuzzy agents is
\begin{equation}
\begin{aligned}
\Delta_{\theta_\pi}J(\theta_\pi) = \sum_{\hat{a}_t}Q_\pi(\hat{s}_t,\hat{a}_t)\Delta_{\theta_\pi}\pi(\hat{a}_t|\hat{s}_t;\theta_\pi),
\label{eq16}
\end{aligned}
\end{equation}

Then the global action value $Q_\pi(\hat{s}_t,\hat{a}_t)$ is calculated by the joint critic network. Correspondingly, the mean-squared Bellman error function of joint critic network is
\begin{equation}
\begin{aligned}
L(\theta_{Q_{\pi}}) = \mathbb{E}[(Q_\pi(\hat{s}_t,\hat{a}_t)-y_t)^2]
\label{eq17}
\end{aligned}
\end{equation}
with the global target $y_t = \hat{r}_t + \gamma Q_\pi(\hat{s}_{t+1},\hat{a}_{t+1}|_{\hat{a}_{t+1}\sim\pi(\hat{s}_{t+1})})$.

Finally, in order to ensure that the target network tends to be stable in the iterative process, the soft update is carried out with the update rate $\tau\ll 1$. The $target$ actor network is ${\theta_{\pi'}}\leftarrow\tau{\theta_{\pi'}}+(1-\tau)\theta_{\pi}$
and the $target$ critic network is ${\theta_{Q_{\pi'}}} \leftarrow \tau{\theta_{Q_{\pi'}}} + (1-\tau)\theta_{Q_\pi}$.

\subsection{FL-CTDE for Maximizing SE of CF XL-MIMO}
The processing flow of the FL-CTCE architecture is shown in Fig. \ref{fig2}(b), all the agents are deployed at the UEs. Hence, all the UEs independently complete the action allocation based on the local information, while the CPU uniformly completes the policy update based on the global information. Compared with the FL-CTCE, the FL-CTDE based on the multi-agent deep deterministic policy gradient (MADDPG) to optimize power allocation coefficients.

With the architecture of the FL-CTDE, each fuzzy agent calculates its own policy gradient of the local actor network according to the joint abstract observation and action. Also, the objective function for the $i$-th $\pi_i$ can be designed as $L(\pi_i) = \sum_{\hat{s}_{i,t}}p_{\pi}(\hat{s}_{i,t})\sum_{\hat{a}_{i,t}}\pi(\hat{a}_{i,t}|\hat{s}_{i,t})\hat{r}_{i,t}$.

Correspondingly, the $i$-th fuzzy reward $\hat{r}_{i,t}$ is based on the global fuzzy action $\hat{a}_t$ and observation $\hat{s}_t$, leading to a centralized global action value $Q_\pi(\hat{s}_t,\hat{a}_t)$, which is calculated by the $i$-th critic network. The policy gradient of local actor network for $\pi_i$ is
\begin{equation}
\begin{aligned}
\Delta_{\theta_{\pi_i}}J(\theta_{\pi_i}) = \sum_{\hat{a}_{i,t}}Q_\pi(\hat{s}_t,\hat{a}_t)\Delta_{\theta_{\pi_i}}\pi_i(\hat{a}_{i,t}|\hat{s}_{i,t};\theta_{\pi_i}).
\label{eq22}
\end{aligned}
\end{equation}

However, different from the FL-CTCE, $\Delta_{\theta_\pi}\pi(\hat{a}_t|\hat{s}_t;\theta_\pi)$ in (\ref{eq16}) is the output by the joint policy network, while $\Delta_{\theta_{\pi_i}}\pi_i(\hat{a}_{i,t}|\hat{s}_{i,t};\theta_{\pi_i})$ in (\ref{eq22}) is the output by the local policy network.
Therefore, the mean-squared Bellman error function of the joint critic network for the $i$-th fuzzy agent is
\begin{equation}
\begin{aligned}
L(\theta_{Q_{\pi}}) = \mathbb{E}[(Q_\pi(\hat{s}_t,\hat{a}_t)-y_{i,t})^2]
\label{eq23}
\end{aligned}
\end{equation}
with the local target $\!y_{i,t} = \hat{r}_{i,t} + \gamma Q_\pi(\hat{s}_{t+1},\hat{a}_{t+1}|_{\hat{a}_{t+1}\sim\pi(\hat{s}_{t+1})})\!$.

Similar to the FL-CTCE, soft update is carried out in combination with the
current network. The $target$ actor network is ${\theta_{\pi_i'}}\leftarrow\tau{\theta_{\pi_i'}}+(1-\tau)\theta_{\pi_i}$
and the $target$ critic network is ${\theta_{Q_{\pi'}}} \leftarrow \tau{\theta_{Q_{\pi'}}} + (1-\tau)\theta_{Q_\pi}$. Both the procedure of the FL-CTCE and the FL-CTDE for maximizing SE performance are summarized in Algorithm 1.

\section{Numerical Results}
We consider a CF XL-MIMO system in an $1\times1$ $\text{km}^2$ with a warp-around scheme \cite{[13]}.
The LSF coefficient is computed by
$\beta_{m,k}[\text{dB}]=-30.5-36.7\log_{10}\left(d_{mk}/1\text{m}\right)+F_{mk}$, where $d_{mk}$ is the distance between BS $m$ and UE $k$ (taking the 10 m height difference into account), and $F_{mk}\sim \mathcal{N}(0,4^2)$ is the shadow fading. Besides, we set up the experimental environment and complete the simulation with PyTorch, and the training works are executed with an Nvidia GeForce GTX 3060 Graphics Processing Unit.

\subsection{Comparison of Total SE}
\begin{figure}[t]  
\centering
\subfigure[L-MMSE combining.]{\includegraphics[scale=0.5]{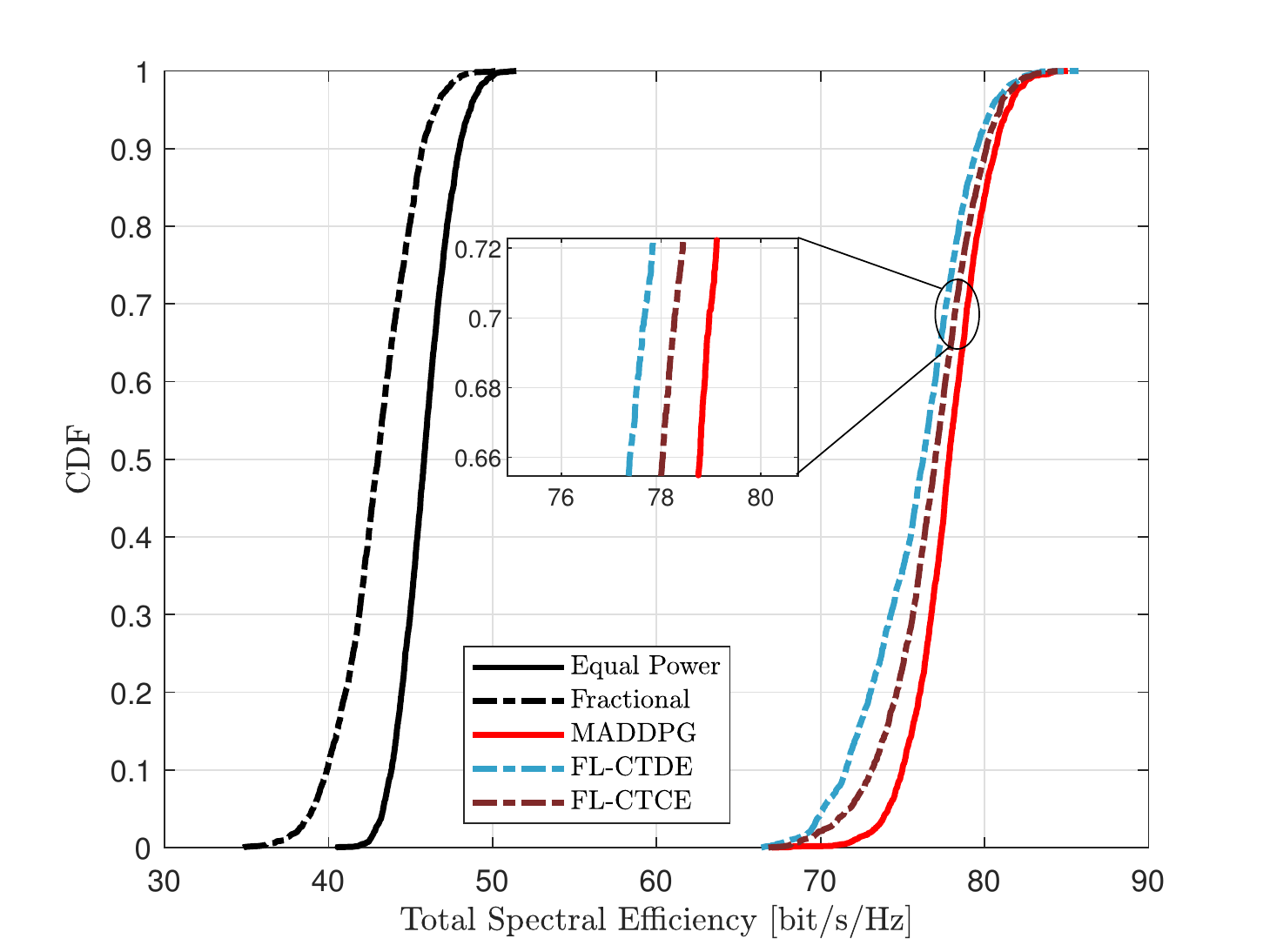}}\\
\hspace{0.195in}
\subfigure[MR combining.]{\includegraphics[scale=0.5]{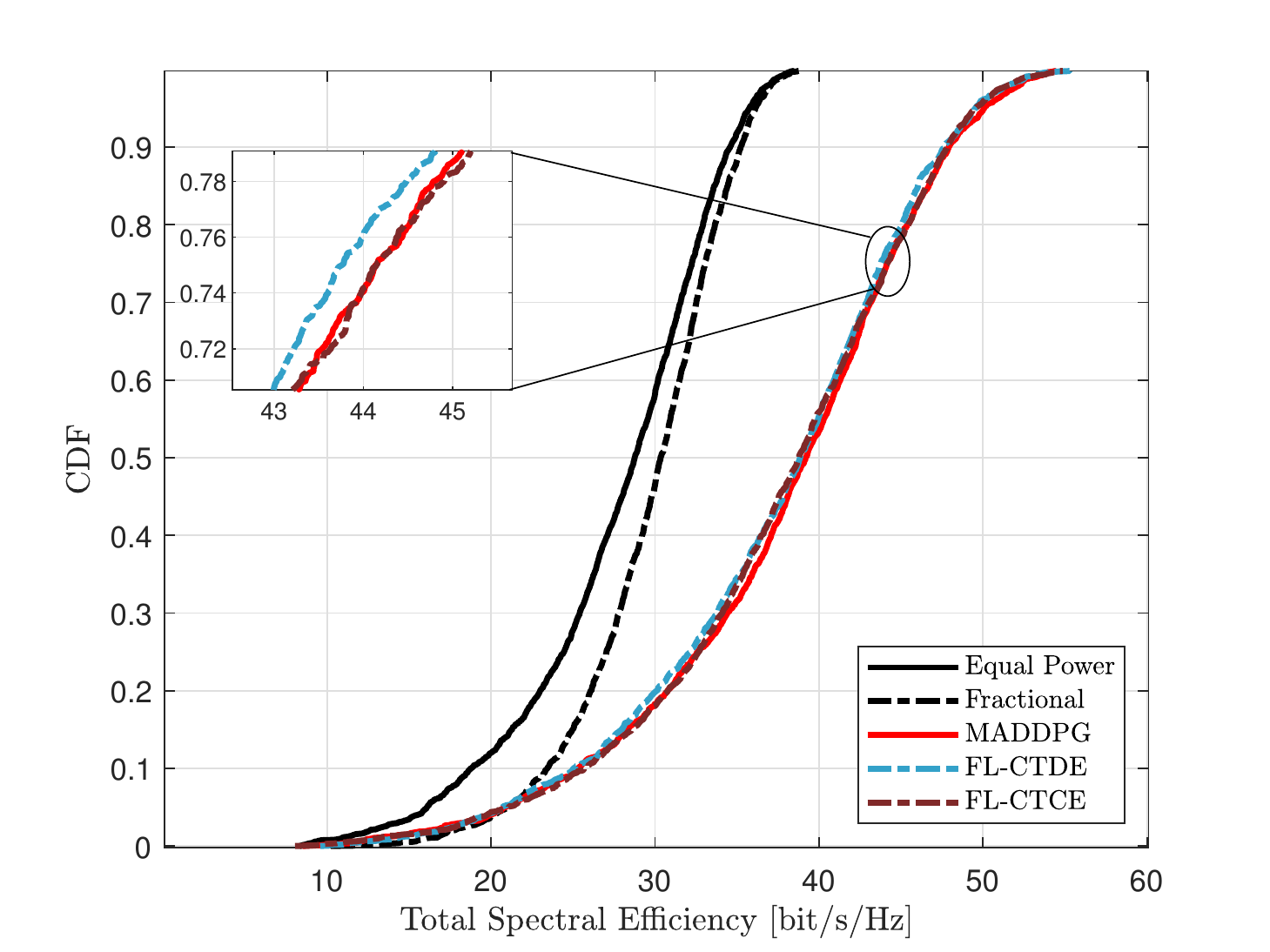}}  
\caption{CDF of total SE for L-MMSE and MR combining with $M=9$, $K=6$, $N_r=N_{H_r}\times N_{V_r}=81$, and $N_s=N_{H_s}\times N_{V_s}=9$.}  
\label{fig3}
\end{figure}
We firstly investigate the effects of different power control schemes on the system performance. Fig. \ref{fig3} shows the cumulative distribution function (CDF) of total SE with L-MMSE and MR combining with $M=9$, $K=6$, $N_r=N_{H_r}\times N_{V_r}=81$, $N_s=N_{H_s}\times N_{V_s}=9$, and $\Delta_s = \Delta_r = \lambda/3$, respectively. For L-MMSE combining shown in Fig. \ref{fig3}(a), we observe that the three MARL-based methods undoubtedly outperform other conventional optimization-based methods since they are based on the reasonable power control in the iterative process, which can suppress the potential inter-user interference. As for MR combining, compared with Fig. \ref{fig3}(a), the proposed methods even outperform the conventional MARL-based method in terms of the SE performance. This is because the lower limit function of policy is designed in the fuzzy system to avoid allocating abnormal power coefficients in the training process. However, we notice that all the schemes suffer from a large SE performance loss, which is caused by the inability of MR combining to effectively suppress the interference. Moreover, since the FL-CTCE updates the policy network based on the global state and action, which makes its performance always better than the FL-CTDE.

\subsection{Comparison of Power Consumption}
\begin{figure}[t]
    \centering
    \includegraphics[scale=0.53]{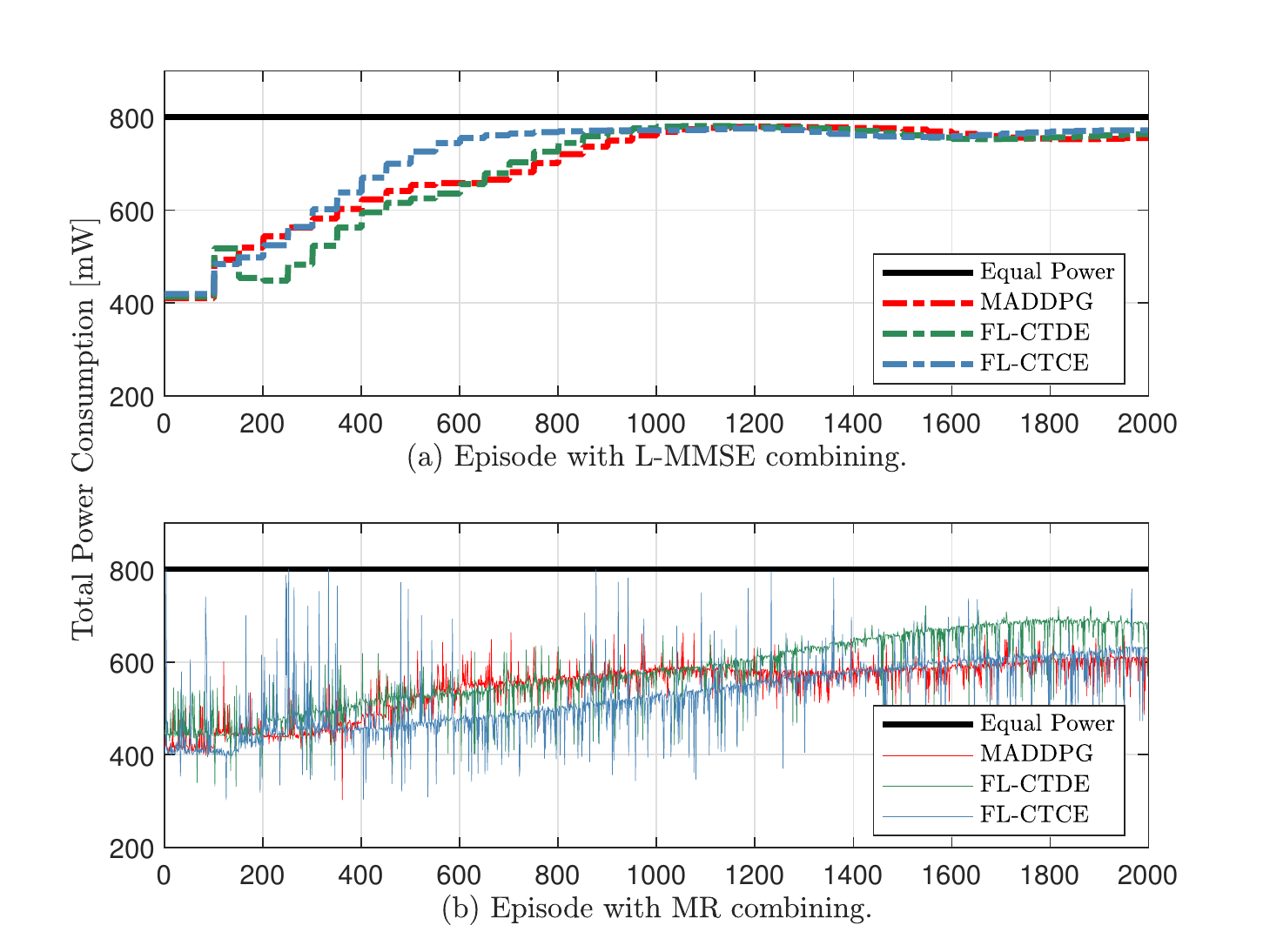}
    \caption{Power consumption for L-MMSE and MR combining with $M=9$, $K=6$, $N_r=N_{H_r}\times N_{V_r}=81$, and $N_s=N_{H_s}\times N_{V_s}=9$.}
    \label{fig4}
\end{figure}
This subsection investigates the power consumption of the proposed schemes presented earlier. Fig. \ref{fig4} depicts the training curve of the power consumption with L-MMSE and MR combining, we can observe that the power consumption slowly approaches to a stable value with the number of training episodes increased. Moreover, by comparing both the L-MMSE and MR combining, we notice that the former is more effective in restraining the potential inter-user interference. In this case, the policy network is almost not affected by the randomly deployed UEs, such that the output power information under the same policy network will eventually become consistent. Therefore, the power consumption of the L-MMSE combining in each small training cycle always tends to a constant value, while the output power consumption with MR combining always fluctuates.

\subsection{Comparison of Computational Complexity}
In Table \ref{tab1}, we present the average run-time, for 2000 episodes, of the proposed methods and MARL-based method with L-MMSE and MR combining.
It is clear that the FL-CTCE and the FL-CTDE can reduce the computational complexity. Thanks to the merit of parallel computing, the FL-CTDE requires the least computational complexity. Besides, we observe that the three MARL-based methods with L-MMSE combining utilize complex combination vectors to reduce the inter-user interference and the computational complexity is nearly three times than that of MR combining.
\begin{table}[htbp]
\caption{Computational Complexity For L-MMSE And MR Combining}
\begin{center}
\begin{tabular}{|c|c|c|c|}
\hline
{\small{{Algorithm}}} & \small{L-MMSE Combining} \normalsize{[s]}& \small{MR Combining} \normalsize{[s]}\\
\hline
\small{{FL-CTCE}}& \small{1.124} & \small{0.382} \\
\hline
\small{{FL-CTDE}}& \small{1.056} & \small{0.347} \\
\hline
\small{{MADDPG}}& \small{1.182} & \small{0.411} \\
\hline
\end{tabular}
\label{tab1}
\end{center}
\end{table}
\section{Conclusion}
In this paper, we investigated the uplink SE maximization of CF XL-MIMO system through power control. Two MARL-based methods with fuzzy logic, i.e., FL-CTCE and FL-CTDE approaches, were proposed. The FL-CTCE delegates most computation burden to the CPU for centralized processing, which is mainly applicable to the situations with limited capacity of terminal equipment. By contrast, the FL-CTDE exploits parallel computing to reduce computing time, which is more suitable for large networks. Our results showed that the proposed methods leveraging fuzzy logic can effectively reduce the computational complexity, enjoying better realizability in practical application scenarios than the conventional MARL-based algorithms. In the future work, we will focus on the downlink power control problem for CF XL-MIMO with the proposed methods of FL-CTCE and FL-CTDE.
\bibliographystyle{IEEEtran}
\bibliography{ref_xl_new}
\end{document}